\documentclass{article}
\usepackage{amsmath}
\usepackage{amssymb}

\def\demo{%
  \par\topsep6pt plus6pt
  \trivlist
  \item[\hskip\labelsep\it Proof.]\ignorespaces}
\def\enddemo{\qed \endtrivlist}
\expandafter\let\csname enddemo*\endcsname=\enddemo

\def\qedsymbol{\ifmmode\bgroup\else$\bgroup\aftergroup$\fi
  \vcenter{\hrule\hbox{\vrule
height.6em\kern.6em\vrule}\hrule}\egroup}
\def\qed{\ifmmode\else\unskip\nobreak\fi\quad\qedsymbol}

 \newtheorem{thm}{Theorem}[section]
 
 \newtheorem{lem}{Lemma}[section]
 
 \newtheorem{exm}{Example}[section]
 \newtheorem{rem}{Remark}[section]
 
 \newtheorem{alg}{Algorithm}[section]

\newcommand{\ra}[1]{\textrm{rank}({#1})}  
\newcommand{\ind}[1]{\textrm{ind}({#1})}  

\font\ssr=cmss8

\font\sst=cmtt8

\font\ssi=cmti8

\title{\bf Computing generalized inverses using LU factorization of matrix product}
\begin{footnotesize}
\author{\frenchspacing
\bf Predrag S. Stanimirovi\' c, Milan B. Tasi\' c\footnote{Corresponding author\ }\\
{\ssi University of Ni\v{s}, Department of Mathematics, Faculty of Science,}\\
{\ssi Vi\v segradska 33, 18000 Ni\v s, Serbia} \\
{\ssi E-mail:} {\sst pecko@pmf.ni.ac.yu},\ \ {\sst milan12t@ptt.yu}\ \\
}
\end{footnotesize}
\date{}
\begin{document}

\maketitle

\begin{abstract}
An algorithm for computing $\{2,3\},\{2,4\}$, $\{1,2,3\},\{1,2,4\}$-inverses and
the Moore-Penrose inverse of a given rational matrix $A$ is established.
Classes $A\{2,3\}_s$ and $A\{2,4\}_s$ are characterized
in terms of matrix products $(R^*A)^{\dagger}R^*$ and $T^*(AT^*)^{\dagger}$,
where $R$ and $T$ are rational matrices with appropriate dimensions and corresponding rank.
The proposed algorithm is based on these general representations and the Cholesky factorization
of symmetric positive matrices. The algorithm is implemented in
programming languages {\ssr MATHEMATICA} and {\ssr DELPHI} and illustrated via examples.
Numerical results of the algorithm, corresponding to the Moore-Penrose inverse,
are compared with corresponding results obtained by several known methods for computing
the Moore-Penrose inverse.

\frenchspacing \itemsep=-1pt
\begin{description}
\item[] AMS Subj. Class.: 15A09, 68Q40.
\item[] Key words: Cholesky factorizations, Generalized inverses, Moore-Penrose inverse, {\ssr MATHEMATICA\/}, {\ssr DELPHI\/}.
\end{description}
\end{abstract}

\section{Introduction}
Let $\mathbf {C}$ be the set of complex numbers, $\mathbf {C}^{m \times n}$ be the set of $m \times n$ complex matrices,
and $\mathbf {C}^{m \times n}_r$ is a subset of $\mathbf {C}^{m \times n}$ consisting matrices
of rank $r$: $\mathbf {C}^{m \times n}_r\!=\!\{X\in \mathbf {C}^{m \times n}\, |\,\,\, \ra{X}\!=\!r\}$.
As usual, $\mathbf {C}(x)$ denotes the set of rational
functions with complex coefficients in the variable $x$. The set of
$m\times n$ matrices with elements belonging to $\mathbf{C}(x)$ is denoted by
$\mathbf {C}(x)^{m\times n}$.
By $I_r$ and $I$ we denote the identity matrix of the order $r$, and identity
matrix of an appropriate order, respectively. By $\mathbf{O}$ is denoted an appropriate null matrix.

\smallskip
For any matrix $A$ of the order $m\times n$ consider the following matrix
equations in $X$, where $*$ denotes conjugate and transpose:
$$(1)\quad AXA\! =\! A \quad (2) \quad XAX\! =\! X \quad (3) \quad
(AX)^*\! =\! AX \quad (4) \quad (XA)^*\! =\! XA. $$
In the case $m=n$ we also consider equations
$$(5) \quad AX=XA\quad \quad (1^k)\quad A^{k+1}X=A^k .$$
For a sequence ${\mathcal S}$ of elements from the set
$\{1,2,3,4,5,1^k\}$, the set of matrices obeying the equations
with corresponding indicative numbers contained in ${\mathcal S}$ is denoted by $A\{{\mathcal S}\}$.
A matrix from $A\{{\mathcal S}\}$ is called an ${\mathcal S}$-inverse of $A$.
The matrix $X=A^\dagger $ is said to be the
Moore-Penrose inverse of $A$ satisfies equations (1)--(4).
The group inverse $A^\#$ is the unique $\{1,2,5\}$ inverse of $A$, and exists
if and only if $\ind{A}=\min\limits _k \{k:\,\ra{A^{k+1}}=\ra{A^k}\}=1$.
A matrix $X=A^D$ is said to be the Drazin inverse of $A$ if
$(1^k)$ (for some positive integer $k$), $(2)$ and $(5)$ are satisfied. In the case $\ind{A}=1$, the Drazin inverse of $A$ is
equal to the group inverse of $A$. If $A$ is nonsingular, it is easily seen that $\ind{A}=0$ and $A^D=A^{-1}$.

\smallskip
The rank of generalized inverse $X$ is important, and it will be convenient to consider the subset $A\{i,j,k\}_s$ of
$A\{i,j,k\}$, consisting $\{i,j,k\}$-inverses of rank $s$ (see \cite{Ben}).

\smallskip
In the literature are known various methods for computing the Moore-Penrose inverse (see for example \cite{Ben}, \cite{WWQ}).
The most commonly implemented method in programming languages is the Singular Value Decomposition (SVD) method,
that is implemented, for example, in the
"pinv" function from Matlab, as well as in the standard {\ssr MATHEMATICA} function "PseudoInverse" \cite{Wol}.
This method is very accurate, but time consuming when the matrix is large. Other well-known methods are
Greville's algorithm, the full rank QR factorization by Gram-Schmidt orthonormalization (GSO), and iterative
methods of various orders \cite{Ben}. A number of expansions of the Moore-Penrose inverse can also be used to
develop direct methods \cite{Rakha}, \cite{Stanimirovic}.

\smallskip
A class of direct methods for computing pseudoinverses is derived from the {\it full-rank factorization} $A=PQ$
of $m\times n$ matrix $A$ of rank $r$, where $P$ is $m\times r$, $Q$ is $r\times n$, and $P,Q$ are both of rank $r$.
These methods are investigated in many papers (see for example \cite{Ben, Rao, Stanimirovic, WWQ}).
After the full-rank factorization, we have the general representation of the Moore-Penrose inverse
$A^{\dagger}=Q^{\dagger}P^{\dagger}$, where $Q^{\dagger}=Q^*(QQ^*)^{-1},\/P^{\dagger}=(P^*P)^{-1}P^*.$
General representations for various classes of $\{2\}$-inverses and the Drazin inverse are obtained in \cite{Stanimirovic}.

\smallskip
Chen et all derived a deterministic iterative algorithm for
computing the Moore-Penrose inverse and rank of matrix $A\in \mathbf
{C}^{m \times n}$ in \cite{Chen}. This algorithm is called {\it
successive matrix powering} and it is based on successive squaring
of a composite matrix $T\!\!=\!\!\left[ \begin{array}{cc} P & Q \\
\mathbf{O} & I \end{array}\right]$, where $P\!\!=\!\!(I-\beta A^*
A), Q\!\!=\!\!\beta A^*$ and $\beta $ is a relaxation parameter. Wei
established {\it successive squaring algorithm} to approximate the
Drazin inverse in \cite{YWei}. The Drazin inverse is expressed in
the form of successive squaring of the composite matrix
$T\!\!=\!\!\left[ \begin{array}{cc} P & Q \\ \mathbf{O} & I
\end{array}\right]$, where $P\!\!=\!\!(I-\beta A^{k+1}), Q\!=\!\beta
A^k$.

\smallskip
In the paper \cite{Courrieu}, Courrieu proposed an algorithm for fast computation of the Moore-Penrose inverse
of real matrices, which is based on known reverse order law (eq. 3.2 from \cite{Rakha}), and on the full rank Cholesky
factorization of possibly singular, symmetric positive matrices (Theorem 4 from \cite{Courrieu1}).

\smallskip
In the present paper we use the {\it LU}-factorization from \cite{Courrieu1}. An arbitrary
matrix $A$ has an LU-factorization if it can be expressed
as the product $A=LU$ of a lower-triangular matrix $L$ and an upper
triangular matrix $U$. When it is possible, we say that $A$
has an {\it LU-decomposition}. It turns out that this factorization
(when exists) is not unique. If $L$ has 1's on it's main diagonal,
then it is called a {\it Doolittle factorization}.  If $U$ has 1's
on its main diagonal, then it is called a {\it Crout factorization}. When
$L=U^*$, it is called the {\it Cholesky decomposition}.
In each of these cases, the following is valid:
$$A^{\dagger}=U^{\dagger}L^{\dagger}=U^*(UU^*)^{-1}(L^*L)^{-1}L^*.$$

An implementation of the Cholesky factorizations in {\ssr MATHEMATICA} can be found on the web site
$$http://math.fullerton.edu/mathews/n2003/CholeskyBib.html.$$

This paper is a generalization of the paper \cite{Courrieu} to sets of $\{2,3\}, \{2,4\}$-inverses
and to the set of rational matrices.

\smallskip
Many numerical algorithms for computing the Moore-Penrose inverse
lack numerical stability. Also, when rounding error is present, we have to identify
some small quantity as being zero.
Moreover, it is well-known that the Moore-Penrose inverse is not
necessarily a continuous function of the elements of the matrix. The existence of this
discontinuity is an additional problem in the pseudoinverse computation.
It is clear that cumulative round off errors should be totally eliminated.
This is possible only by symbolic computation. During the symbolic implementation,
variables are stored in the "exact" form or can be left "unassigned" (without numerical values),
resulting in no loss of accuracy during the calculation \cite{Kar}.

\smallskip
Algorithms for computing generalized inverses of polynomial and/or rational matrices are so far based upon the
Leverrier-Faddeev algorithm and the Grevile's algorithm.
Computation of the Moore-Penrose inverse of polynomial and/or rational matrices which uses the
Leverrier-Faddeev algorithm is investigated in \cite{Fra,Jon,Kar,Kar1,Tze}.
An algorithm of the Leverrier-Faddeev type for computing the Moore-Penrose inverse
of a polynomial matrix is introduced in the paper \cite{Kar}.
Implementation of this algorithm, in the symbolic computational
language {\ssr MAPLE\/}, is described in \cite{Jon}.
Furthermore, in \cite{Jon} it is described an implementation of the algorithm
for computing the Moore-Penrose inverse of a singular rational matrix.

\smallskip
A representation and corresponding algorithm
for computing the Drazin inverse of a singular one-variable polynomial matrix of arbitrary degree
are introduced in \cite{Ji}, \cite{Pecko1}.
Corresponding algorithm for two-variable polynomial matrix and its implementation is introduced in \cite{Wei}.
Also, an effective version of given algorithm is established in the paper \cite{Wei}.

\smallskip
A general finite algorithm for computing various classes of
generalized inverses of a polynomial matrix is introduced in \cite{Pecko2}.
This algorithm is based on the Leverrier-Faddeev algorithm.

Computation of the Moore-Penrose inverse of one-variable polynomial and/or rational matrices,
arising from the Grevile's algorithm, is introduced in \cite{Pecko}.
Corresponding two-dimensional case is investigated in \cite{Stanimirovic3}.

\smallskip
The Moore-Penrose inverse is used in the evaluation of the least square solution of linear system $Ax=b$,
even with rank deficient matrices \cite{Ben}.
In fact, the Moore-Penrose inverse $A^\dagger $ is
defined as that matrix which, when postmultiplied by $b$, yields the minimum-length
least-square solution $x$ of the possibly inconsistent equation $Ax \approx b$, for any $b$.
Also, the Moore-Penrose inverse can have
valuable applications in neurocomputational learning procedures \cite{Courrieu}.
Moreover, in the literature it is known a number of applications of generalized inverses of polynomial matrices
\cite{Jon,Kar,Kar1,Lov,Lov1,Lov2}.

\smallskip
This paper is a first attempt to compute $\{i,j,k\}$ generalized inverses of one-variable rational matrices
using the method from \cite{Courrieu}.

\smallskip
In the second section we characterize classes $A\{2,3\}_s$, $A\{2,4\}_s$, $A\{1,2,3\}$ and $A\{1,2,4\}$
in terms of matrix products $(R^*A)^{\dagger}R^*$ and $T^*(AT^*)^{\dagger}$,
where $R$ and $T$ are rational matrices with appropriate dimensions and corresponding rank.
Using these representations, we introduce a method for computing $\{i,j,k\}$-inverses of prescribed rank $s$ of a given
rational matrix $A$.
When $A$ is a constant matrix, in two partial cases ($R=A$ or $T=A$), we get
an algorithm for computing the Moore-Penrose inverse, alternative with corresponding one introduced in \cite{Courrieu}.

Algorithm introduced in this paper is implemented in programming package {\ssr MATHEMATICA}, and it is applicable
to rational and constant matrices.
Corresponding algorithm, applicable only to constant matrices, is also implemented in
the programming language {\ssr DELPHI}.
Symbolic implementation in {\ssr MATHEMATICA} is illustrated via examples in Section 3.
We especially consider the partial case of the implementation, which computes the Moore-Penrose inverse of a constant matrix.
This partial case of the implementation is compared with several known methods for computing the Moore-Penrose inverse.

\section{Representations of \{i,j,k\} inverses for rational matrices} \setcounter{equation}{0}

In the following lemma we modify known representations for $\{2,3\}$, $\{2,4\}$-inverses of prescribed rank,
introduced in \cite{Ben}.
We also extend these representations, known for complex matrices, to the set of one-variable rational matrices.

\begin{lem} Let $A\in \mathbf{C}(x)^{m\times n}_r$ and $0<s\leq r$, $m_1,n_1\geq s$
be chosen integers. Then the following general representations for pseudoinverses are valid:

\smallskip \smallskip
{\bf (a)\/}
$A\{2,4\}_s=\left\{ (YA)^\dagger Y |\ Y\in \mathbf{C}(x)^{n_1\times m},\ YA\in \mathbf{C}(x)^{n_1\times n}_s \right\}$.

\smallskip
{\bf (b)\/}
$A\{2,3\}_s=\left\{ Z(AZ)^\dagger |\ Z\in \mathbf{C}(x)^{n\times m_1},\ AZ\in \mathbf{C}(x)^{m\times m_1}_s \right\}$.
\end{lem}

\demo
{\bf (a)} The inclusion
$A\{2,4\}_s \!\supseteq \! \left\{ (YA)^\dagger Y |\ Y\!\in \!\mathbf{C}(x)^{n_1\times m},\ YA\!\in \!\mathbf{C}(x)^{n_1\times n}_s \right\}$
can be proved in a similar way as in \cite{Ben}.

\smallskip
To prove the opposite inclusion, choose an arbitrary $X\in A\{2,4\}_s$.
Consider a full-rank factorization $X\!=\!FG$, $F\!\in \!\mathbf{C}(x)^{n\times s}_s, G\!\in \!\mathbf{C}(x)^{s\times m}_s$.
Since $X\!\in \!A\{2\}$, we get
$$FGAFG=FG$$
or
$$F(GAF-I_s)G=\mathbf{O}.$$
This implies
$$GAF=I_s.$$
Now, it is not difficult to verify $F\in (GA)\{1,2,3,4\}$, or equivalently $F=(GA)^\dagger$.
Consequently,
$$X=(GA)^\dagger G\in \left\{ (YA)^\dagger Y |\ Y\in \mathbf{C}(x)^{s\times m},\ YA\in \mathbf{C}(x)^{s\times n}_s \right\}.$$
Using
$$\left\{ (YA)^\dagger Y | Y\!\in \!\mathbf{C}(x)^{s\times m}, YA\!\in \!\mathbf{C}(x)^{s\times n}_s \right\}\!\subseteq \!
\left\{ (YA)^\dagger Y | Y\!\in \!\mathbf{C}(x)^{n_1\times m}, YA\!\in \!\mathbf{C}(x)^{n_1\times n}_s \right\}$$
we prove part {\bf (a)}.

\smallskip
Part {\bf (b)} can be verified in a similar way.
\enddemo

\begin{rem}
In the case $m_1=n_1=s$, in the case of constant matrices, we get an improvement in the proof of Theorem 6 and Theorem 7
from $\cite{Ben}$ $($p. $63)$.
\end{rem}

Analogous representations of $\{1,2,3\}$ and $\{1,2,4\}$-inverses we derive in the case
$s=r=\ra{A}$.

\begin{lem} Let $A\in \mathbf{C}(x)^{m\times n}_r$ and $m_1,n_1\geq r$
be chosen integers. Then the following statements are valid for
the sets $A\{1,2,4\}$, $A\{1,2,3\}$ and the Moore-Penrose inverse:

\smallskip \smallskip
{\bf (a)\/} $A\{1,2,4\}=\left\{ (YA)^\dagger Y|\ Y\in \mathbf{C}(x)^{n_1\times m},\
YA\in \mathbf {C}(x)^{n_1\times n}_r \right\} .$

\smallskip
{\bf (b)\/} $A\{1,2,3\}=\left\{ Z(AZ)^\dagger |\ Z\in \mathbf{C}(x)^{n\times m_1},\
AZ\in \mathbf {C}(x)^{m\times m_1}_r \right\} .$

{\bf (c)\/} $A^\dagger =(A^*A)^\dagger A^*=A^*(AA^*)^\dagger.$
\end{lem}

\smallskip
Now we are in a position to propose the next theorem for computing
$\{2,3\}$, $\{2,4\}$ inverses of prescribed rank as well as $\{1,2,3\}$ and $\{1,2,4\}$ inverses of a given
matrix $A\in \mathbf{C}(x)^{m\times n}_r $. This theorem is a customization of Lemma 2.1
to generalized LU factorization from \cite{Courrieu} and \cite{Courrieu1}.

\begin{thm} Consider rational matrix $A\!\in \!\mathbf {C}(x)^{m \times n}_r$.
Let $0\!<\!s\!\leq \!r$ be randomly chosen integer and assume that $m_1,n_1$ are positive integers satisfying $m_1,n_1\!\geq \!s$.
Then the following statements are valid:

\smallskip
{\bf (a)\/}
\begin{equation}
A\{2,4\}_s\!=\! \left\{ L(L^*L)^{-2} L^*(R^*A)^*R^*|\ R\! \in\! \mathbf {C}(x)^{m\times n_1}_s, R^*A\! \in \! \mathbf {C}(x)^{n_1\times n}_s\right\},
\end{equation}
where $(R^*A)^*(R^*A)=LL^*$ is the Cholesky factorization and $L^*$ is without the zero rows.

\smallskip
{\bf (b)\/}
\begin{equation}
A\{2,3\}_s\! =\! \left\{ T^*(AT^*)^*L (L^*L)^{-2} L^*|\ T\! \in\!  \mathbf {C}(x)^{m_1\times n}_s, AT^*\! \in \! \mathbf {C}(x)^{m\times m_1}_s \right\},
\end{equation}
where $(AT^*)(AT^*)^*=LL^{*}$ is the Cholesky factorization and $L^*$ is without the zero rows.

\smallskip
{\bf (c)\/}
\begin{equation}
A\{1,2,4\}\! =\! \left\{ L(L^*L)^{-2}L^*(R^*A)^*R^*|\ R\! \in \! \mathbf {C}(x)^{m \times n_1}_r, R^*A\! \in \! \mathbf {C}(x)^{n_1\times n}_r\right\},
\end{equation}
where $(R^*A)^*(R^*A)=LL^*$ is the Cholesky factorization and $L^*$ is without the zero rows.

\smallskip
{\bf (d)\/}
\begin{equation}
A\{1,2,3\}\! =\! \left\{ T^*(AT^*)^{*}L(L^*L)^{-2}L^*|\ T\! \in \! \mathbf {C}(x)^{m_1\times n}_r, AT^*\! \in \! \mathbf {C}(x)^{m\times m_1}_r\right\},
\end{equation}
where $(AT^*)(AT^*)^*=LL^*$ is the Cholesky factorization and $L^*$ is without the zero rows.

\smallskip
{\bf (e)\/}
\begin{equation}
A^\dagger\! =\! L(L^*L)^{-2}L^*(A^*A)^*A^*,
\end{equation}
where $(A^*A)^*(A^*A)=LL^*$ is the Cholesky factorization and $L^*$ is without the zero rows,
or
\begin{equation}
A^{\dagger}= A^*(AA^*)^*L(L^*L)^{-2}L^*,
\end{equation}
where $(AA^*)(AA^*)^*=LL^*$ is the Cholesky factorization and $L^*$ is without the zero rows.
\end{thm}

\demo {\bf (a)}
Various expressions for computing the Moore-Penrose inverse of the matrix product $(AB)^{\dagger}$ are considered in \cite{Rakha}.
We use the following:
\begin{equation}
  (AB)^{\dagger} =B^*(A^*ABB^*)^{\dagger}A^*\label{jedan}.
\end{equation}

Applying (\ref{jedan}) in the case $A=R^*A$, $B=I$, the
Moore-Penrose inverse $(R^*A)^{\dagger}$ can be found as

\begin{equation}
  (R^*A)^{\dagger} =((R^*A)^*(R^*A))^{\dagger}(R^*A)^*.
\end{equation}

There is an unique upper triangular matrix $S$ with exactly $n-s$
zero rows, such that $S^*S=(R^*A)^*(R^*A)$, where the
computation of $S$ is an application of the extension of the usual Cholesky
factorization from \cite{Courrieu}, \cite{Courrieu1} on matrix $(R^*A)^*(R^*A)$. Removing
the zero rows from $S$, one obtains an $r\times n$ matrix of rank $r$, denoted by $L^*$.
The following is evident:
\begin{equation}
(R^*A)^*(R^*A)=S^*S=LL^*.
\end{equation}
Applying (2.9) in (2.8), we get

\begin{equation}
(R^*A)^{\dagger}=(LL^*)^{\dagger}(R^*A)^*.
\end{equation}
Applying now (2.7) in the case $A=L$, $B=L^*$, one can verify the following
\begin{equation}
  (LL^*)^{\dagger}= L(L^*L)^{-1}(L^*L)^{-1}L^*.
\end{equation}
Multiplying  $(R^*A)^{\dagger}$ by $R^*$ from the right, in view of (2.10) and (2.11) we obtain
$$(R^*A)^{\dagger}R^*=L(L^*L)^{-2}L^*(R^*A)^*R^*.$$
Now, the proof follows from Lemma 2.1, part {\bf (a)}.

\smallskip
{\bf (b)\/} This part of theorem can be proved in a similar way as part
{\bf (a)}, applying part {\bf (b)} from Lemma 2.1 and $A=I$, $B=AT^*$.
Also, in this case $m$ and $m_1$ appears instead of $n$ and $n_1$, respectively.

\smallskip
Parts {\bf (c)}, {\bf (d)} and {\bf (e)} can be proved applying Lemma 2.2.
\enddemo

Using Theorem 2.1, we now state the following algorithm which generates classes $A\{2,4\}_s$ and $A\{2,3\}_s$.

\begin{alg}{\label{alg1}} Choose $m\times n$ rational matrix $A$ and consider randomly chosen $m_1\times n_1$ rational matrix $R$,
where $m_1=m$ and $n_1$ is arbitrary integer $\geq r$, or $n_1=n$ and $m_1$ is arbitrary integer $\geq r$.

\smallskip
\item[{\it Step 1.\/}] If $n=n_1$ then compute $G:=(AR^*)(AR^*)^*$ and set $n=m$ and logical variable $trans=True$;

\noindent else compute $G:=(R^*A)^*(R^*A)$.

\smallskip
\item[{\it Step 2.\/}] Find Cholesky factorization of matrix $G\!=\!LL^*$ and drop zero rows from $L^*$.

\smallskip
\item[{\it Step 3.\/}] If $trans$ then return $R^*(AR^*)^*L(L^*L)^{-2}L^*$;

\quad else return $L(L^*L)^{-2}L^*(R^*A)^*R^*$.
\end{alg}

This algorithm is applicable to class of rational matrices if we implement them in symbolic
programming languages like {\ssr MATHEMATICA}, {\ssr MAPLE} etc.
Our implementation is developed in {\ssr MATHEMATICA}.
However, because of the problems with the simplification in rational expressions,
this algorithm is not convenient for the implementation in high level
programming languages such as {\ssr C++, DELPHI, VISUAL BASIC} etc.
Therefore, our implementation in language {\ssr DELPHI} is applicable only for constant matrices.

\section{Examples}

\begin{exm} In this example we consider constant matrices.
Let $A\!\in \!\!\mathbf {C}^{6 \times 4}_4$ and $R\!\in \!\!\mathbf {C}^{6 \times 6}_4$ be the following matrices:
\begin{small}
$$A =\left(
  \begin{array}{cccc}
  -1&0&1&2\\-1&3&0&-1\\10&-1&1&3\\0&1&-1&-3\\1&-1&0&1\\1&0&-1&-2\\
\end{array}
\right), \quad R=\left(\begin{array}{cccccc}
3&-1&3&1&2&-1\\0&-1&0&0&-2&1\\3&1&-3&1&2&-1\\0&-1&0&0&-2&1\\3&1&3&-1&2&-1\\0&-1&0&0&-2&1
\end{array}
\right).$$
\end{small}

\noindent Applying the function {\tt ModGinvCholesky[A,R]}, described in Appendix, we obtain
\begin{small}
$$L=\left(
  \begin{array}{cccccc}
    2\sqrt{627} & 0 & 0 & 0\\
    -92 {\sqrt{\frac{3}{209}}} & 2 {\sqrt{\frac{6819}{209}}} & 0 & 0 \\
    70 {\sqrt{\frac{3}{209}}} & -2632 {\sqrt{\frac{3}{475057}}} & 4 {\sqrt{\frac{1409}{2273}}} & 0\\
    \frac{634}{{\sqrt{627}}} & \frac{-24758}{{\sqrt{1425171}}} & \frac{12778}{{\sqrt{3202657}}} & 10 {\sqrt{\frac{26}{1409}}}\\
  \end{array}
\right)$$ and $$A\{1,2,4\}=\left(
  \begin{array}{cccccc}- \frac{1}{10}  &0&\frac{1}{10}&0&- \frac{1}{10}
 &0\\ 1&\frac{1}{2}&0&\frac{1}{2}&1&\frac{1}{2}\\
- \frac{13}{10}  &-1&\frac{3}{10}&-1&-
\frac{43}{10}  &-1\\
\frac{11}{10}&\frac{1}{2}&- \frac{1}{10}
&\frac{1}{2}&\frac{21}{10}&\frac{1}{2}\end{array} \right).$$
\end{small}
Let us mention that conditions of Theorem 2.1, part {\bf (c)} are valid.
\end{exm}

\begin{exm} Let us consider matrix $A$ of rank 3:

\begin{small}
{\tt A=\{\{x+1,x,5\},\{x+2,x,3\},\{x-1,x,1\},\{x+3,x,2\}\}. }
\end{small}

\smallskip
Choose the following matrix $R$ of rank 2:

\smallskip
\begin{small}
{\tt R=\{\{x+1,2\},\{x+1,2\},\{x+1,3\},\{x+1,3\}\}.}
\end{small}

\smallskip
\noindent In accordance with part {\bf (a)} of Theorem 2.1, function {\tt ModGinvCholesky[A,R]}
generates the following $\{2,4\}$-inverse of $A$ of rank 2:

\smallskip
\begin{small}
\noindent$\begin{array}{llllll}
In[3]:=ModGinvCholesky[A, R]\\
Out[3]=\{ \{ \frac{-21 - 30\,x + 4\,x^2}{49 + 140\,x + 204\,x^2},
\frac{-21 - 30\,x + 4\,x^2}{49 + 140\,x + 204\,x^2},
\frac{56 + 80\,x - 4\,x^2}{49 + 140\,x + 204\,x^2},
\frac{56 + 80\,x - 4\,x^2}{49 + 140\,x + 204\,x^2}\}, \\
\hspace{1.5cm}\{ \frac{-2\,x\,\left( 17 + 2\,x \right) }{49 + 140\,x + 204\,x^2},
\frac{-2\,x\,\left( 17 + 2\,x \right) }{49 + 140\,x + 204\,x^2},\frac{2\,x\,\left( 43 + 2\,x \right) }{49 + 140\,x + 204\,x^2},
\frac{2\,x\,\left( 43 + 2\,x \right) }{49 + 140\,x + 204\,x^2}\} ,\\
\hspace{1.5cm}\{ \frac{14 + 34\,x + 40\,x^2}{49 + 140\,x + 204\,x^2},\frac{14 + 34\,x + 40\,x^2}{49 + 140\,x + 204\,x^2},
- \frac{21 + 44\,x + 40\,x^2}{49 + 140\,x + 204\,x^2}  ,- \frac{21 + 44\,x + 40\,x^2}{49 + 140\,x + 204\,x^2} \} \}
\end{array}$
\end{small}
\end{exm}

\smallskip
\begin{exm} In this example we choose matrices $A$ and $T$ satisfying conditions
imposed in part {\bf (d)} of Theorem 2.1. Then an $\{1,2,3\}$-inverse is generated in the output:

\smallskip
\begin{small}
\noindent$\begin{array}{lll}
In[4]:=A=\{\{1+x,x,5,2+x,x,3\},\{-1+x,x,1,3+ x,x,2\},\\
\hspace{2cm}\{-2+x,x,1,3+x,x,2\},\{-3+x,-1+x,1,1+x,x,1\}\};\\
In[5]:=T=\{\{1+x,2,2+x,1,-1+x,3\},\{2+x,3,3+x,1,-2+x,2\},\\
\hspace{2cm}\{3+x,3,3+x,-1,-2+x,1\},\{2+x,3,3+x,4,-1+x,1\},\\
\hspace{2cm}\{2+x,3,3+x,-1,-1+x,1\}\};\\
In[6]:=ModGinvCholesky[A, T]
\end{array}$

\smallskip
\noindent$\begin{array}{llllll}
Out[6]\!\!=\!\!
\{ \{ 0,1,-1,0\} ,\{ \frac{5031 - 13465\,x + 14101\,x^2 - 130\,x^3 - 975\,x^4}{30186 - 78744\,x + 63024\,x^2 + 49340\,x^3 + 7800\,x^4},\\
   \frac{-70434 + 89855\,x - 4908\,x^2 + 18453\,x^3 + 6615\,x^4}{30186 - 78744\,x + 63024\,x^2 + 49340\,x^3 + 7800\,x^4},
    \frac{-75465 + 82542\,x + 12803\,x^2 + 22101\,x^3 + 6180\,x^4}{30186 - 78744\,x + 63024\,x^2 + 49340\,x^3 + 7800\,x^4} ,\\
   \frac{-25155 + 20168\,x + 11555\,x^2 + 4153\,x^3 + 540\,x^4}{30186 - 78744\,x + 63024\,x^2 + 49340\,x^3 + 7800\,x^4}\} ,
  \{ \frac{5031 - 11455\,x + 7726\,x^2 + 5905\,x^3 + 975\,x^4}{30186 - 78744\,x + 63024\,x^2 + 49340\,x^3 + 7800\,x^4},\\
   \frac{-70434 + 90587\,x + 28674\,x^2 - 3784\,x^3 - 1695\,x^4}{30186 - 78744\,x + 63024\,x^2 + 49340\,x^3 + 7800\,x^4},
  \frac{75465 - 85296\,x - 45272\,x^2 - 4707\,x^3 + 540\,x^4}{30186 - 78744\,x + 63024\,x^2 + 49340\,x^3 + 7800\,x^4},\\
   \frac{-25155 + 22250\,x + 19052\,x^2 + 4011\,x^3 + 180\,x^4}{30186 - 78744\,x + 63024\,x^2 + 49340\,x^3 + 7800\,x^4}\} ,
  \{ - \frac{5031 - 15463\,x + 15407\,x^2 + 8530\,x^3 + 975\,x^4}{30186 - 78744\,x + 63024\,x^2 + 49340\,x^3 + 7800\,x^4}  ,\\
   \frac{-50310 + 38335\,x + 81884\,x^2 + 21697\,x^3 + 735\,x^4}{30186 - 78744\,x + 63024\,x^2 + 49340\,x^3 + 7800\,x^4},
   \frac{75465 - 86214\,x - 56095\,x^2 + 1091\,x^3 + 2780\,x^4}{30186 - 78744\,x + 63024\,x^2 + 49340\,x^3 + 7800\,x^4},\\
   \frac{-35217 + 49192\,x + 543\,x^2 - 12483\,x^3 - 2540\,x^4}{30186 - 78744\,x + 63024\,x^2 + 49340\,x^3 + 7800\,x^4}\} ,
  \{ \frac{-12 + 3071\,x - 13389\,x^2 + 4760\,x^3 + 1950\,x^4}{30186 - 78744\,x + 63024\,x^2 + 49340\,x^3 + 7800\,x^4},\\
   \frac{-96960 + 246044\,x + 119747\,x^2 - 42630\,x^3 - 15150\,x^4}{30186 - 78744\,x + 63024\,x^2 + 49340\,x^3 + 7800\,x^4},
   \frac{106974 - 261537\,x - 153182\,x^2 + 21230\,x^3 + 11200\,x^4}{30186 - 78744\,x + 63024\,x^2 + 49340\,x^3 + 7800\,x^4},\\
   \frac{-29838 + 61289\,x + 78394\,x^2 + 22290\,x^3 + 2000\,x^4}{30186 - 78744\,x + 63024\,x^2 + 49340\,x^3 + 7800\,x^4}\} ,
 \{ \frac{5031 - 17461\,x + 16713\,x^2 + 17190\,x^3 + 2925\,x^4}{30186 - 78744\,x + 63024\,x^2 + 49340\,x^3 + 7800\,x^4},\\
   - \frac{-140868 + 87781\,x + 221884\,x^2 + 111187\,x^3 + 15885\,x^4}{30186 - 78744\,x + 63024\,x^2 + 49340\,x^3 + 7800\,x^4},
   \frac{-166023 + 97482\,x + 251041\,x^2 + 118599\,x^3 + 16220\,x^4}{30186 - 78744\,x + 63024\,x^2 + 49340\,x^3 + 7800\,x^4},\\
   - \frac{-65403 + 39808\,x + 75665\,x^2 + 28527\,x^3 + 3260\,x^4}{30186 - 78744\,x + 63024\,x^2 + 49340\,x^3 + 7800\,x^4}  \} \}
\end{array}$
\end{small}
\end{exm}

\begin{exm} In this example we generate $\{1,2,4\}$-inverse using the following matrices $A$ and $R$:

\smallskip
\begin{small}
{\tt A=\{\{x+1,x,5\},\{x+2,x,3\},\{x-1,x,1\},\{x+3,x,2\},\{x-2,x,1\},\{x+3,x,2\}\}.}

\smallskip
{\tt R=\{\{1+x,2,2+x,1,-1+x\},\{2+x,3,3+x,1,-2+x\},\{3+x,3,3+x,-1,-2+x\},}

\qquad {\tt \{2+x,3,3+x,4,-1+x\},\{2+x,3,3+x,-1,-1+x\},\{1+x,2,2+x,1,-1+x\}\}.}
\end{small}

\medskip
\begin{small}
\noindent$\begin{array}{lllllll}
In[9]:=ModGinvCholesky[A, R]\\
Out[9]\!\!=\!\!
\{\{ \frac{1596 - 2292\,x + 2542\,x^2}{-41601 + 39942\,x - 27634\,x^2},\frac{-5593 + 2166\,x + 4766\,x^2}{83202 - 79884\,x + 55268\,x^2},
   \frac{2310 - 5520\,x + 1282\,x^2}{-41601 + 39942\,x - 27634\,x^2},\\
\frac{13573 - 13626\,x + 7944\,x^2}{41601 - 39942\,x + 27634\,x^2},
   \frac{10549 - 4878\,x + 7922\,x^2}{-83202 + 79884\,x - 55268\,x^2},\frac{1596 - 2292\,x + 2542\,x^2}{-41601 + 39942\,x - 27634\,x^2}\} ,\\
  \{ \frac{23961 - 36414\,x + 38983\,x^2 - 5084\,x^3}{-83202\,x + 79884\,x^2 - 55268\,x^3},
   \frac{-38171 + 3108\,x + 76654\,x^2 - 9532\,x^3}{4\,x\,\left( 41601 - 39942\,x + 27634\,x^2 \right) },
 \frac{60564 - 14028\,x + 34011\,x^2 + 2564\,x^3}{83202\,x - 79884\,x^2 + 55268\,x^3},\\
   \frac{74774 - 105294\,x + 62993\,x^2 - 15888\,x^3}{83202\,x - 79884\,x^2 + 55268\,x^3},
  \frac{29743 - 69888\,x - 4194\,x^2 + 15844\,x^3}{166404\,x - 159768\,x^2 + 110536\,x^3},
   \frac{23961 - 36414\,x + 38983\,x^2 - 5084\,x^3}{-83202\,x + 79884\,x^2 - 55268\,x^3}\} ,\\
 \{ \frac{16905 - 21399\,x + 20869\,x^2}{83202 - 79884\,x + 55268\,x^2},
   \frac{-7\,\left( -5257 + 1862\,x + 4414\,x^2 \right) }{4\,\left( 41601 - 39942\,x + 27634\,x^2 \right) },
  \frac{18228 + 3465\,x + 14261\,x^2}{-83202 + 79884\,x - 55268\,x^2},\\
\frac{-7\,\left( 5446 - 5735\,x + 2597\,x^2 \right) }{83202 - 79884\,x + 55268\,x^2},
   \frac{8281 + 25270\,x + 12302\,x^2}{166404 - 159768\,x + 110536\,x^2},\frac{16905 - 21399\,x + 20869\,x^2}{83202 - 79884\,x + 55268\,x^2}\} \}
\end{array}$
\end{small}
\end{exm}

\smallskip
\begin{exm} In this example we choose matrices $A$ and $T$ satisfying conditions
imposed in part {\bf (b)} of Theorem 2.1. Then an $\{2,3\}$-inverse of rank 2 is generated:

\smallskip
\begin{small}
\noindent$\begin{array}{lllll}
In[10]:=A=\{\{x + 1, x, 5\},\{x + 2, x, 3\}, \{x - 1, x, 1\}, \{x + 3, x, 2\}\};\\
In[11]:=T=\{\{x + 1, 2, x - 1\}, \{x + 2, 1, x - 1\}\}\\
In[12]:=ModGinvCholesky[A, T]\\
Out[12]= \{ \{ - \frac{41 - 139\,x + 88\,x^2 + 25\,x^3 + x^4}{329 -
1168\,x + 984\,x^2 + 380\,x^3 + 35\,x^4} ,
   \frac{30 - 99\,x + 83\,x^2 + 31\,x^3 + 3\,x^4}{329 - 1168\,x + 984\,x^2 + 380\,x^3 + 35\,x^4},\\
  - \frac{55 - 239\,x + 222\,x^2 + 97\,x^3 + 9\,x^4}{329 - 1168\,x + 984\,x^2 + 380\,x^3 + 35\,x^4}  ,
   \frac{85 - 290\,x + 228\,x^2 + 82\,x^3 + 7\,x^4}{329 - 1168\,x + 984\,x^2 + 380\,x^3 + 35\,x^4}\} ,
  \{ \frac{-136 + 69\,x + 207\,x^2 + 35\,x^3 + x^4}{329 - 1168\,x + 984\,x^2 + 380\,x^3 + 35\,x^4},\\
   - \frac{69 - 170\,x + 40\,x^2 + 26\,x^3 + 3\,x^4}{329 - 1168\,x + 984\,x^2 + 380\,x^3 + 35\,x^4}  ,
  \frac{-38 + 3\,x + 373\,x^2 + 117\,x^3 + 9\,x^4}{329 - 1168\,x + 984\,x^2 + 380\,x^3 + 35\,x^4},
   - \frac{31 - 254\,x + 246\,x^2 + 82\,x^3 + 7\,x^4}{329 - 1168\,x + 984\,x^2 + 380\,x^3 + 35\,x^4}  \} ,\\
 \{ \frac{59 - 148\,x + 79\,x^2 + 10\,x^3}{329 - 1168\,x + 984\,x^2 + 380\,x^3 + 35\,x^4},
  \frac{13 - 41\,x + 23\,x^2 + 5\,x^3}{329 - 1168\,x + 984\,x^2 + 380\,x^3 + 35\,x^4},
   \frac{31 - 122\,x + 71\,x^2 + 20\,x^3}{329 - 1168\,x + 984\,x^2 + 380\,x^3 + 35\,x^4},\\
   \frac{-18\,{( -1 + x) }^2}{329 - 1168\,x + 984\,x^2 + 380\,x^3 + 35\,x^4}\} \}
\end{array}$
\end{small}
\end{exm}

We compare the processor time conditioned by different algorithms for computing the Moore-Penrose inverse
of constant matrices in the next table.
Test matrices are taken from \cite{Zie}, and considered in the partial case $a=1$.
The test matrix name we state in the first column .
Processor times required by the standard {\ssr MATHEMATICA} function $PseudoInverse[\,]$ (see {\cite{Wol}})
are allocated in the second column of the table.
Results corresponding to function $Partitioning[\,]$ from {\cite{Stanimirovic2}} are placed
in the third column.
Fourth column is filled by the results generated by using the Leverrier-Faddeev algorithm from \cite{Jon}.
Results produced by applying {\ssr MATHEMATICA} implementation of the algorithm from \cite{Courrieu}
are placed in the next column,
and the last two columns are arranged for the {\ssr MATHEMATICA} and {\ssr DELPHI} implementation of Algorithm \ref{alg1}.
We use $R=A$ in {\ssr MATHEMATICA} functions $ModGinvCholesky[\,]$ and {\ssr DELPHI} function $A1234$
to compute the Moore-Penrose inverse.
For matrix dimensions above $20\times 20$ an application of the function $ModGinvCholesky[\,]$ gives the information:
{\it "Result for Inverse of badly conditioned matrix $<<1>>$ may contain significant numerical errors"!}
These cases are marked by the sign '*' in the table.
Also, the sign '-' denotes a long processor time needed for the computation.

\medskip
\begin{footnotesize}
\noindent \begin{tabular}{|l|l|l|l|l|l|l|}
\hline
Test   &Math.        & Math.       & Math.     & Math.  & Math.  & Delphi\\
matrix &PseudoInverse& Partitioning&Lev.Faddeev&Courrieu&Alg. 2.1&Alg. 2.1\\
\hline
S5 & 0.079  & 0.016 & 0.001 & 0.001 & 0.001 & 0.062\\
S10&0.031  & 0.031  & 0.001 & 0.015 & 0.015 & 0.062\\
S25&- & 0.125 & 0.062 & 0.047 & 0.109 * & 0.062\\
S50&- & 1.187 & 2.516 & 0.375 & 0.687 * & 0.940\\
S100&- & 9.204 & 44.375 & 2.297 & 5.781 * & 1.850\\
F5&0.125 & 0.031 & 0.001 & 0.001 & 0.001& 0.047 \\
F10&1.094 & 0.016 & 0.001 & 0.015 & 0.015 & 0.047\\
F25&- & 0.047 & 0.156 & 0.110 & 0.250 * & 0.062\\
F50&- & 0.485 & 2.672 & 0.703 & 2.328 * & 0.940\\
F100&- & 2.812 & 42.844 & 5.782 & 17.594 * & 1.850\\
A5&0.25 & 0.006 & 0.001 & 0.001 & 0.001& 0.047 \\
A10&1.344 & 0.015 & 0.001 & 0.015 & 0.015&0.062\\
A25&- & 0.063 & 0.171 & 0.093 & 0.265 * & 0.062\\
A50&- & 0.484 & 2.766 & 0.766 & 2.218 * & 0.940\\
A100&- & 2.750 & 43.781 & 5.844 & 16.954 * & 1.850\\
\hline
\end{tabular}

\end{footnotesize}

\smallskip
{\centerline{Table 1. Processor time in Seconds for constant matrices}}

\section{Conclusion}

We introduce an algorithm for computing $\{1,2,3\}$, $\{1,2,4\}$-inverses, $\{2,3\}$, $\{2,4\}$-inverses of prescribed rank
as well as for computing the Moore-Penrose inverse for one-variable rational matrices.
Our method uses the representations of $\{i,j,k\}$-inverses based on the matrix product
involving the Moore-Penrose inverse and factors of the full-rank Cholesky factorization from \cite{Courrieu1}.
On the other hand, a large number of representations and algorithms are available for computing generalized inverses of rational
and/or polynomial matrices {\cite{Fra,Ji,Jon,Kar,Kar1,Pecko,Pecko1,Pecko2,Stanimirovic2,Stanimirovic3,Tze,Wei}.
But, generalized inverses in these papers are computed using the
Leverrier-Faddeev algorithm and the Grevile's algorithm.
The algorithm proposed in this paper is an extension of the paper \cite{Courrieu} to various
classes of $\{i,j,k\}$-inverses and to rational matrices.
When the input matrix is constant, in a certain case $R=A$, we get an algorithm for
computing the Moore-Pernose inverse, alternative with respect to the algorithm introduced in \cite{Courrieu}.

\smallskip
Introduced algorithm is implemented in two different programming languages: {\ssr MAT\-HEMATICA} and {\ssr DELPHI}.
The implementation in {\ssr DELPHI} is appropriate only for constant matrices.
In the constant matrix case we compare processor time required by these implementations of Algorithm 2.1
with respect to standard {\ssr MATHEMATICA} function Pseudoinverse, implementation of
Grevile's partitioning method, implementation of Leverrier-Faddeev algorithm
and the {\ssr MATHEMATICA } implementation of the algorithm from \cite{Courrieu}.

\smallskip
Column 2 is a confirmation of the statement that the method used in {\ssr MATHEMATICA} function
{\tt PseudoInverse} is time consuming for large matrices.
The results from columns 3, 4, 5 and 6 in Table 1 again confirm known fact that {\ssr MATHEMATICA} (and other symbolic packages) is not
applicable for large scale test problems.
Our numerical experience shows that the algorithm introduced in \cite{Courrieu} is
superior with respect to the Grevile's partitioning algorithm for test matrices of smaller dimensions.
But, the algorithm from \cite{Courrieu} is inferior with respect to partitioning method
in the case when test matrices of relatively great order from \cite{Zie} are used.
Leverrier-Faddeev algorithm produces the best results for test matrices of small dimensions and the
worst results for test matrices of greater dimensions.

\smallskip
Algorithm 2.1 produces inferior results with respect to
algorithm from \cite{Courrieu} for matrix dimensions greater than $20\times 20$.
The reason is clear. Algorithm from \cite{Courrieu} computes the Moore-Penrose inverse using the Cholesky factorization of
the matrix products $A^*A$ or $AA^*$. On the other side, Algorithm 2.1 factorizes the matrix products $(A^*A)^*(A^*A)$ or $(AA^*)(AA^*)^*$,
which produce bigger numbers causing badly conditioned matrices.
But, our method for computing the Moore-Penrose inverse arises from a general algorithm, which
is limited by the application of symmetric positive matrices $(R^*A)^*(R^*A)$ or $(AT^*)(AT^*)^*$.

\section{APPENDIX}

For the sake of completeness we present the {\ssr MATHEMATICA} and {\ssr DELPHI} code for the
implementation of Algorithm 2.1.

\subsection{Mathematica code}

In the following function we implement the Cholesky factorization.

\begin{footnotesize}\begin{verbatim}
Cholesky[A0_,n_]:=Module[{A=A0,i,k,m,L,U},
   L=Table[0,{n},{n}];
    For[k=1,k<=n,k++,
      L[[k,k]]=Sqrt[A[[k,k]]-Sum[L[[k,m]]^2,{m,1,k-1}]];
      For[i=k+1,i<=n,i++,
        L[[i,k]]=(A[[i,k]]-Sum[L[[i,m]]L[[k,m]],{m,1,k-1}])/L[[k,k]]]];
    U = Transpose[L];
    Return[L] ]
\end{verbatim}\end{footnotesize}

In the auxiliary function {\tt Adop[a,j]} we drop the last $n-j$ columns from the matrix $a$.
This function is used for the elimination of last zero rows in the matrix $Transpose[a]$.

\begin{footnotesize}\begin{verbatim}
Adop[a_List,j_]:=Module[{m, n},
    {m,n}=Dimensions[a];
    Return[Transpose[Drop[Transpose[a],-(n-j)]]];]
\end{verbatim}\end{footnotesize}

Function {\tt GinvCholesky[A]} implements the algorithm from \cite{Courrieu}

\begin{footnotesize}\begin{verbatim}
GinvCholesky[A0_List]:=Module[{m,n,trans,A=A0,L,M,Y},
    {m,n}=Dimensions[A0]; trans=False;
    If[m<n, trans=True; A=A0.Transpose[A0]; n=m,
      A=Transpose[A0].A0];
    L=Cholesky[A,n]; L=Simplify[Adop[L,MatrixRank[A0]]];
    M=Inverse[Transpose[L].L];
    If[trans,Y=Transpose[A0].L.M.M.Transpose[L],
      Y=L.M.M.Transpose[L].Transpose[A0]];
    Return[Simplify[Y]]]
\end{verbatim}\end{footnotesize}

Function {\tt ModGinvCholesky[A,R]} implements Algorithm 2.1.

\begin{footnotesize}\begin{verbatim}
ModGinvCholesky[A_List,R_List]:=
  Module[{m,m1,n1,n,rr,trans=False,L,M,Y,G,G1},
  {m,n}=Dimensions[A]; {m1,n1}=Dimensions[R];
  If[n==n1,trans=True;
      G=Simplify[A.Transpose[R].Transpose[A.Transpose[R]]]; n=m,
      G=Simplify[Transpose[Transpose[R].A].Transpose[R].A]];
  L=Cholesky[G,n];L=Adop[L,Min[MatrixRank[A],MatrixRank[R]]];
  M=Inverse[Transpose[L].L];
  If[trans,Y=Transpose[R].Transpose[A.Transpose[R]].L.M.M.Transpose[L],
      Y=L.M.M.Transpose[L].Transpose[Transpose[R].A].Transpose[R]];
  Return[Simplify[Y]]]
\end{verbatim}\end{footnotesize}

\subsection{Delphi code}

We present the main part of {\ssr DELPHI } code for
computing $A\{i,j,k\}$-inverses of a given constant matrix $A$. Elementary functions used in computations are:
function $TransMat(\/)$ which computes the transpose matrix, function $MatMatR(\/)$ for the matrix multiplication,
function $MatrixRank(\/)$ for computing the matrix rank,
the function which generates the matrix consisting of first $i$ columns of a given matrix, called $FifstIColumns(\/)$,
and the function $InversionM(\/)$ used for the usual matrix inversion. These functions are not restated here.

Cholesky factorization is implemented in the following function.

\begin{footnotesize}\begin{verbatim}
procedure TForm1.Cholesky(A0:matrix;var C0:matrix;n:integer);
var i,j,p,q:integer;s:extended;s1:real;
begin
  For i:=1 to n do
    For j:=1 to n do C0[i,j]:=0;
  For p:=1 to n do
  begin
      s:=0;
      for q:=1 to p-1 do s:=s+C0[p,q]*C0[p,q]
      s1:=A0[p,p]-s;
      if s1<0.00000000001 then s1:=0;
      C0[p,p]:=Sqrt(s1);
      if C0[p,p]<>0 then
      begin
         for i:=p+1 to n do
         begin
            s:=0;
            for j:=1 to p-1 do s:=s+C0[i,j]*C0[p,j];
            C0[i,p]:=(A0[i,p]-s)/C0[p,p];
         end;
      end;
  end;
end;
\end{verbatim}\end{footnotesize}

Function {\tt A1234} implements Algorithm 2.1.

\begin{footnotesize}\begin{verbatim}
procedure TForm1.A1234(A1,R1:matrix;var L:matrix);
var trans:boolean; minrank:integer;
    Y1,L1,L2,G1,G2,G3,G4,G5,G6,G7,G8:matrix;
begin
    trans:=false;
    if n=nn then
      begin
        trans:=true;
        TransMat(R1,mm,nn,G1); MatMatR(A1,G1,m,n,mm,G2);
        TransMat(G2,m,mm,G3);  MatMatR(G2,G3,m,mm,m,Y1);
        n:=m;
      end
      else begin
             TransMat(R1,mm,nn,G1); MatMatR(G1,A1,nn,m,n,G2);
             TransMat(G2,nn,n,G3);  MatMatR(G3,G2,n,nn,n,Y1);
           end;
     Cholesky(Y1,L1,n);
     minrank:=MatrixRank(L1,n);
     firstIColumn(L1,minrank,Y1);
     TransMat(Y1,n,minrank,G1);   MatMatR(G1,L1,minrank,n,n,G2);
     InverseM(G2,n,L2);
     if trans then
       begin
         TransMat(R1,mm,nn,G1); MatMatR(A1,G1,m,nn,m,G2);
         TransMat(G2,m,m,G3);   MatMatR(G1,G3,nn,mm,m,G4);
         MatMatR(G4,L1,nn,m,n,G5); MatMatR(G5,L2,nn,n,n,G6);
         MatMatR(G6,L2,nn,n,n,G7); TransMat(L1,n,n,G8);
         MatMatR(G7,G8,nn,n,n,L);
         WriteY(L,nn,n);
       end
       else begin
              MatMatR(L1,L2,n,n,n,G1);   MatMatR(G1,L2,n,n,n,G2);
              TransMat(L1,n,n,G3);       MatMatR(G2,G3,n,n,n,G4);
              TransMat(R1,mm,nn,G5);     MatMatR(G5,A1,nn,mm,n,G6);
              TransMat(G6,nn,n,G7);      MatMatR(G7,G5,n,nn,mm,G8);
              MatMatR(G4,G8,n,n,mm,L);
              WriteY(L,n,mm);
            end;
end;
\end{verbatim}\end{footnotesize}

\enddocument
\bye